\RequirePackage[2020-02-02]{latexrelease}
\documentclass[twocolumn,english,prb]{revtex4-1}
\pdfoutput=1
\usepackage[T1]{fontenc}
\usepackage[latin9]{inputenc}
\usepackage{amsmath}
\usepackage{amssymb}
\usepackage{graphicx}
\usepackage{babel}
\usepackage{xcolor}
\usepackage{natbib}
\usepackage{listings}
\begin{document}
\title{High-$T_c$ superconducting dome in artificial heterostructures made of nanoscale quantum building blocks}
\author{Antonio Valletta}
\thanks{Corresponding author: antonio.valletta@cnr.it}
\affiliation{Consiglio Nazionale delle Ricerche CNR, Institute for Microelectronics and Microsystems, via del Fosso del Cavaliere 100, 00133 Roma, Italy}
\author{Antonio Bianconi}
\thanks{Corresponding author: antonio.bianconi@ricmass.eu}
\affiliation{Rome International Center for Materials Science, Superstripes RICMASS Via dei Sabelli 119A, 00185 Roma, Italy}
\affiliation{Consiglio Nazionale delle Ricerche CNR, Istituto di Cristallografia, via Salaria Km 29.300, Monterotondo (Roma) I-00016, Italy}
\author{Andrea Perali}
\affiliation{School of Pharmacy, Physics Unit, University of Camerino, 62032 Camerino, Italy}
\thanks{Corresponding author: andrea.perali@unicam.it}
\affiliation{Rome International Center for Materials Science, Superstripes RICMASS Via dei Sabelli 119A, 00185 Roma, Italy}
\thanks{Corresponding author: andrea.perali@unicam.it}
\author{Gennady  Logvenov}
\affiliation{Max Planck Institute for Solid State Research, Heisenbergstrasse 1, 70569 Stuttgart, Germany}
\thanks{Corresponding author: g.logvenov@fkf.mpg.de}
\author{Gaetano Campi}
\affiliation{Consiglio Nazionale delle Ricerche, Istituto di Cristallografia IC-CNR, via Salaria Km 29. 300, Monterotondo (Roma) I-00016, Italy}

\begin{abstract}

While the search of high-$T_c$ superconductivity was driven mostly by trial and error methodology searching for novel materials, here we provide a quasi-first-principle quantum theory for engineering superconductivity in artificial high-$T_c$ superlattices (AHTS) with period $d$, ranging from 5.28 down to 3 nanometers, made of superconducting quantum wells of variable thickness $L$.  An important feature of our quantum design is the key role of the interface internal electric field giving Rashba spin-orbit coupling (SOC) in the nanoscale quantum superconducting building blocks. By tuning the geometrical conformational parameter $L/d$ around its magic ratio 2/3 we predict the superconducting dome of $T_c$ versus doping characteristic of unconventional superconductors. Quantum size effects, controlled by $L/d$, change the energy width and splitting of two quantum subbands formed by the electronic space charge confined in superconducting nano-layers. The theoretical superconducting dome $T_c$ versus charge density controlled by the Fano-Feshbach resonance between two superconducting gaps has been able to predict experimental results on cuprate AHTS by tuning the geometry of superlattices of quantum wells made of superconducting layers (S) of thickness $L$ of modulation doped stoichiometric Mott insulator $La_2CuO_4$ with no chemical dopants, with interface space charge confined within normal metal (N) overdoped cuprate layers.

\end{abstract}

 \date{\today}   

\maketitle

\section{Introduction}
Nano-science advances have been recognised by the Nobel prize 2023 to quantum dots research [\onlinecite{manna2023bright}]  
driven by the ability to manipulate and engineer nano-structured materials (NsM) with a microstructure 
characteristic length scale on the order of a few nanometers, typically $1-10$.
While in the old standard paradigm of solid state physics the macroscopic properties have been usually assumed 
to be determined by the homogeneous long range structure of their atomic elements or molecular constituents,
now the focus of the research is addressed to quantum NsM where novel unconventional functionalities emerge in the macroscopic world 
when their building-blocks are in the nanometer regime. 
The functionalities of these quantum ensembles may be tuned by controlling the quantum size effects of the nano-objects used as building blocks.
Nanotechnology therefore offers unique venues to drive quantum science from micro to macro world.
The rational production and engineering of the functional hetero-interfaces between atomically thin nanoscale 
layered units in NsM is especially true in two-dimensional (2D) semiconductor materials [\onlinecite{smith1990theory}]. As a result, 2D materials and 
their heterostructures have been recently modified to function as catalysts, photodetectors, chemical sensors,
memory, logic devices, and single photon emitters [\onlinecite{kahn2020functional}].

\begin{figure}
	\centering
	\includegraphics[scale=0.37]{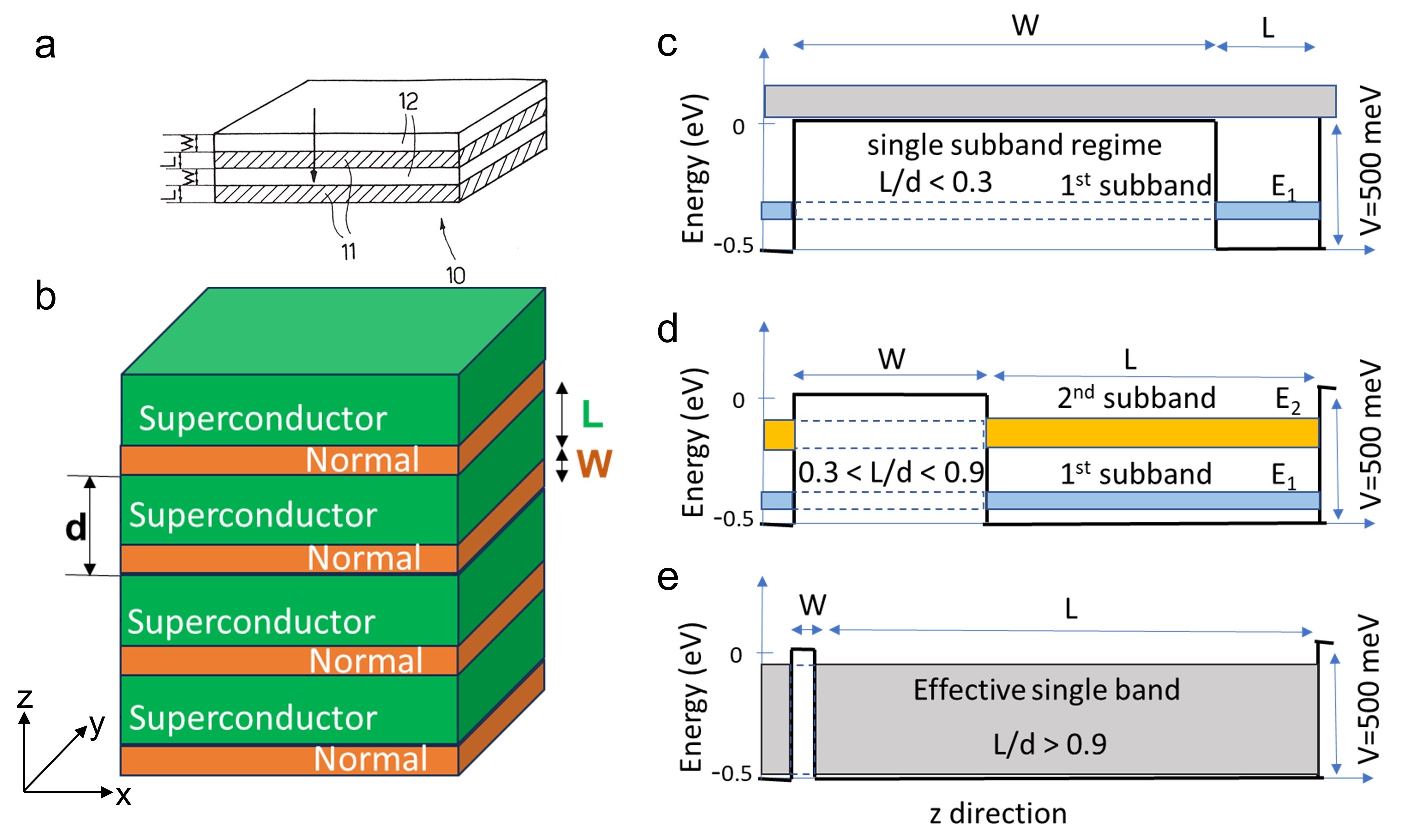} 
	\caption{Panel (a), the heterostructure at the atomic limit made of superconducting units (11) and normal metal units (10) according with the 1993 patent [\onlinecite{bianconi1998european,bianconi2001process}]; Panel (b) shows the realization of a superlattice of quantum wells made of a superconductor, the modulation doped stoichiometric $La_2CuO_4$  of thickness $L$, intercalated by a normal metal, overdoped $La_{1.55}Sr_{0.45}O_4$ of thickness W with period $d=L+W$[\onlinecite{logvenov2023superconducting}]. Panels (c,d,e) show the confinement of the interface space charge, within the potential well of thickness L, which generates the first subband at $E_1$  and the second subband at $E_2$, where $L/d$ of the superlattice is $0.3<L/d<0.9$  for $d=3nm$, panel (d). For $L/d<0.3$ the second subband is pushed above the potential barrier, panel (c), while above $L/d>0.9$, panel (e), the two subbands merge in a single band leaving only one resonant tunnelling subband. The Fano-Feshbach resonance appears in the range of two-gap superconductivity $0.3<L/d<0.9$ shown in panel (d).}
\label{fig:1}
\end{figure}

Recently it has been shown the feasibility of engineering the superconducting critical temperature of a superlattice of quantum wells by tuning the nanoscale geometry of artificial high-$T_c$ superlattices  (AHTS) [\onlinecite{logvenov2023superconducting}] predicted by quantum design [\onlinecite{mazziotti2022spin,mazziottimultigap}].
This work has shown that the macroscopic superconducting properties of NsM can advantageously be engineered by the controlled assembly of nano-objects as nanometer thick layers building-blocks following the claims of the international patent [\onlinecite{bianconi1998european,bianconi2001process}] for the synthesis of
artificial superconductors made of superlattices of quantum wells shown in panel (a) of Fig.\ref{fig:1}, with priority date Dec 7, 1993.  The idea to realise practical artificial  heterostructures at the atomic limit as in natural doped perovskites was presented in ref.[\onlinecite{bianconi1994possibility}] and at the invited talk [\onlinecite{ bianconi1994instability}] given at the 4-th International Conference "Materials and Mechanisms of Superconductivity High-Temperature Superconductors" (M2S-HTSC IV) 
 held in Grenoble (France), 5-9 July 1994. In this work it was proposed that the universal superconducting dome of cuprates, where $T_C$ is plotted versus doping at the insulator-to-metal transition  [\onlinecite{takagi1989superconductor}], is determined by the nanoscale phase separation of insulator and metal nanoscale units. The chemical nanoscale phase separation has been clearly observed in 40 K superconducting oxygen doped $La_2CuO_{4+y}$, between a first oxygen depleted phase $La_2CuO_{4}$ and a second  oxygen rich $La_2CuO_{4.125}$ overdoped phase [\onlinecite{fratini2010scale,poccia2011evolution}], forming the superconducting phase in the space charge interface of the oxygen depleted phase [\onlinecite{jarlborg2013fermi}].
 
These experiments have been confirmed in natural superconducting perovskites using local and fast methods probing the nanoscale structure in these last 38 years, which have shown that natural high-$T_c$ superconductor perovskites are nano-structured materials (NsM) with nanoscale spatial heterogeneity pointing to the emergence of quantum macroscopic functionality at high-temperature due to nanoscale lattice heterogeneity
[\onlinecite{bianconi1996determination,egami2006nano,fratini2010scale,poccia2011evolution,ricci2011nanoscale,campi2015inhomogeneity}]
driven by an arrested nanoscale phase separation 
[\onlinecite{muller1993phase,sigmund2012phase,goodenough1997new,kugel2008model,kagan2021electronic}].

In this scenario quantum size effects appears near the Lifshitz electronic topological transition (ETT). The appearing Fermi surface 
with an extended van Hove singularity triggers the Fano-Feshbach resonance between
first electron pairs in a first large Fermi surface and second electron pairs in the second small Fermi surface with vanishing  group velocity.

Tremendous advances have been reported on the synthesis and device integration of 2D electronic gas at
 oxide interfaces [\onlinecite{baiutti2015high,suyolcu2020design,zhou2023emerging,bonmassar2023superconductivity}]
opening new venues to control and manipulate interfacial Rashba interaction, giving spin-orbit coupling (SOC) in 
atomically thin metallic heterostructure [\onlinecite{krishnia2023large,kong2021tunable,lin2019interface}], where the coexistence 
of quasi-two-dimensional superconductivity and tunable Kondo lattice in superconducting 
heterointerfaces has been observed [\onlinecite{shen2022coexistence,yang2022engineered,helmes2008kondo,li2021display}].

It is known that the two-dimensional electron gases at oxide interfaces can be doped by charge-transfer-induced in modulation-doped hetero-junctions [\onlinecite{dingle1978electron,son2011heterojunction,chen2015extreme,mondal2023modulation,yunoki2007electron}] and recently it 
has been shown that  the superconductivity in atomically thin interface layers  made of a doped Mott insulator can be controlled by quantum geometry confinement 
providing  a new perspective to solve the puzzle of the mechanism driving the formation of the superconducting dome [\onlinecite{logvenov2023superconducting}].

For decades the dominant theoretical research $paradigm$ was based on the assumption of a strongly correlated or a strongly electron-phonon interacting electron liquid in a  homogeneous long range crystalline lattice structure. On the contrary few heretic scientists proposed hat the mystery can be unlocked by the falsification of the  old $paradigm$ with a T.S. Kuhn scientific revolution to a  new $paradigm$ focusing on Quantum Complex Matter called with the nick name $superstripes$ landscape  [\onlinecite{bianconi2020superconductivity}] where quantum size effects in the nano building blocks (layers, stripes or rods) play the key role in the high $T_c$ mechanism in nano-structured  materials. In these last 30 years the Bianconi-Perali-Valletta (BPV) theory has been developed considering nano-building blocks with size of the order 
of 1-5 nm, [\onlinecite{bianconi1994possibility,perali1996gap,valletta1997electronic,bianconi1998superconductivity}] which has been recently
supported by the synthesis driven by the BPV theory of new artificial high $T_c$ superlattices
of quantum wells [\onlinecite{logvenov2023superconducting}]. 

\begin{figure}
	\centering
	\includegraphics[scale=1.2]{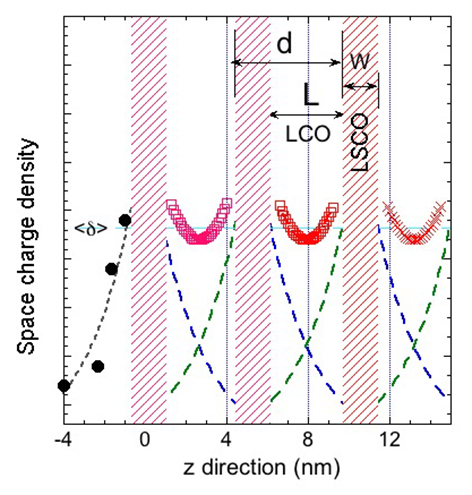} 
	\caption{Space charge (open red icircles) distribution in a nanoscale superlattice made of a superconductor, a modulation doped stoichiometric $La_2CuO_4$  (LCO) of thickness $L$, intercalated by a normal metal, overdoped $La_{1.55}Sr_{0.45}O_4$ (LSCO) of thickness W, with period $d=L+W$ = 3.96 nm and $L/d=$2/3 [\onlinecite{logvenov2023superconducting}]. The interface space charge is confined within the potential well of thickness L for $z>0$.  The exponential decay of the interface space charge outside LSCO measured by Baiutti et al. [\onlinecite{baiutti2015high}] is plotted for $z<0$.}
\label{fig:1a}
\end{figure}

The quantum design of high temperature superconductors made of superlattices of quantum wells, or quantum stripes, where first nanoscale
thick superconducting units  are intercalated by normal metal nanoscale units  has been provided in the patents [\onlinecite{bianconi1998european,bianconi2001process}] for  the $T_c$-resonant amplification via Fano-Feshbach resonance between 
multiple quantum subbands generated by quantum confinement in the superconducting nanoscale units.
The  possibility  of   $T_c$ amplification by Fano resonance between open and close pairing channels in artificial superlattices 
showing two-gap superconductivity was confirmed by the discovery of naturally  layered two-gap superconductors like doped magnesium diborides in 
2001, iron-based perovskite superconductors in 2008 described by the BPV 
theory [\onlinecite{innocenti2010resonant,innocenti2010shape,bianconi2014shape}].
In this work we focus on the BPV theory including SOC  [\onlinecite{mazziotti2022spin}] of artificial superlattices of first superconducting building blocks, atomically thin layers of 
a Mott insulator cuprate perovskite $La_2CuO_4$ doped by interface space charge [\onlinecite{baiutti2015high}] of thickness $L$ intercalated 
 by layers of a normal metal of $La_{1.55}Sr_{0.45}CuO_4$ thickness W with experimental period $d$ in the range between $2.97\ nm$ 
and $5.28\ nm$ [\onlinecite{logvenov2023superconducting}] as shown in Fig.\ref{fig:1a}.
It was already well established that a 2D superconducting layer of thickness 2.6 $nm$, about two lattice units, is formed
at the interface of $La_{1.55}Sr_{0.45}CuO_4$ in the $La_2CuO_4$ side [\onlinecite{suyolcu2020design,ju2022emergence}], hosting 
a superconducting phase of a strongly correlated electron gas [\onlinecite{misawa2016self,tadano2019ab}].
These works have inspired the material design of different classes of artificial nanoscale superlattices where the electronic properties 
of the interface superconductivity can be modified not only by a  careful selection of the materials but also by the thickness of quantum 
units within the stack [\onlinecite{mazziotti2022spin}][\onlinecite{mazziottimultigap}].

The Bianconi, Perali and Valletta (BPV) theory [\onlinecite{bianconi1998superconductivity,perali1996gap,valletta1997electronic}] 
proposes a two-gap superconductivity scenario where a first BCS pairing scattering channel (open channel) 
in a first Fermi surface with a high Fermi energy resonates with a second pairing channel (closed channel) in the BCS-BEC crossover regime
in the second Fermi surface with low Fermi energy in the range of the energy cut-off of the pairing interaction.
[\onlinecite{perali2004bcs,perali2004quantitative,perali2012anomalous,shanenko2012atypical,chen2012superconducting,cariglia2016shape,guidini2014band,guidini2016bcs,doria2016multigap,salasnich2019screening,shanenko2006size,valentinis2016bcs,vargas2020crossband,ochi2022resonant,kagan2019fermi}].
 It has been shown that the highest $T_c$ occurs where
 the chemical potential is tuned by pressure or charge density at the topological Lifshitz transition for the appearing of the second Fermi surface.
Therefore the resonant $T_c$ amplification giving the superconducting dome is a function of the charge density or the  Lifshitz parameter $\eta$ defined as
\begin{equation}
\label{eq:7.1}
\eta=\frac{\mu-E_L}{\omega_0},
\end{equation}
where $\mu$ is the chemical potential, $E_L$ is the second subband bottom energy, and $\omega_0$ is the energy cut-off of the pairing interaction taken here to be $\omega_0$=60 meV. Above the Lifshitz transition for  appearance of a new Fermi surface,  $T_c$  as a function of $\eta$ shows a  quantum resonance between open and closed scattering channels called Fano resonances in multigap systems.
[\onlinecite{fano1961effects}]. 
The BPV theory is based on the calculation of the energy dependent 
pair-transfer exchange interaction in the Bogoliubov-BCS gap equation 
[\onlinecite{bogoliubov1958new}] [\onlinecite{svistunov2015superfluid}].
The energy dependent pair-transfer exchange term is calculated from the overlap of the four wave-functions giving the pair transfer 
between the two Fermi surfaces  
[\onlinecite{perali2004bcs,perali2004quantitative,perali2012anomalous,innocenti2010resonant,innocenti2010shape,innocenti2013isotope,bianconi2013shape,bianconi2015lifshitz,bianconi2015superconductivity,bianconi2020superconductivity,valletta1997electronic,valletta1997t,mazziotti2017possible}]
 solving the Dirac equation [\onlinecite{mazziottimultigap}] for the particular nanoscale superlattice of quantum wells.
 
The $T_c$ amplification as function of energy shows the typical Fano line-shape resonance with a dip 
at the Fano antiresonance near $\eta$=0 followed by a resonant amplification of $T_c$ with the asymmetric line-shape typical of a Fano resonance [\onlinecite{fano1961effects}].
In  superlattices of quantum wells the topological Lifshitz transition from a closed 3D Fermi surface to an open
corrugated cylindrical Fermi surface, called opening a neck, with only a singular critical point in the k-space,
which is not enough to produce a peak in the density of states.
Further amplification of the critical temperature 
$T_c$ for superconductivity at a shape resonance is driven by including spin-orbit-coupling terms  [\onlinecite{mazziottimultigap}] associated with the internal electric field at the interface between the superconducting-normal heterojunction.
In  the presence of Rashba spin-orbit-coupling (SOC)  the $T_c$ shows a maximum at the topological Lifshitz transition from a three dimensional $torus$ to a 
corrugated $cylinder$ associated with an unconventional higher order van Hove singularity (VHS) associated with a circular line of critical points 
and a sharp peak in the DOS increasing with SOC strength [\onlinecite{mazziottimultigap}].  

We have studied the superconducting dome of $T_c$ as a function  of (i) the Lifshitz parameter measuring the position of the chemical potential relative to the Lifshitz transition and (ii) the strength of the electron-phonon interaction in the upper subband in [\onlinecite{mazziotti2022spin}]. In the BPV theory the Fano-Feshbach resonance driven by quantum configuration interaction between a first gap in the BCS regime, resonating with a second gap in the BCS-BEC crossover regime which gives the superconducting dome where  the maximum $T_c$ increases with SOC strength, where we have studied the case of a superlattice of period $d=3$ nm made of  alternating superconducting (S) layers of thickness $ L = 2.3$ nm   and normal (N) metal layers of thickness $W = 0.7$ nm, with a potential barrier $V=500$ meV  made of the normal metal layers intercalated between the superconducting quantum wells with a potential barrier V due to the chemical potential shift at the interface between the overdoped $La_{1.55}Sr_{0.45}CuO_4$ and undoped  $La_2CuO_4$ layers has be taken from the experiments on the chemical shift of doped correlated copper oxides by Fujimori group [\onlinecite{ino1997chemical,fujimori1998chemical,fujimori2002core,yagi2006chemical}]. 
We have shown [\onlinecite{mazziotti2022spin}] that using a weak electron phonon coupling $g=0.3$, in the second subband and $g=0.1$ in the first subband with the energy cut off 60 meV for the pairing interaction $\omega_0$  and spin orbit interaction $\alpha$=0.5, we obtain a superconducting critical temperature maximum around $T_c=40 K$ in AHTS-like natural optimum doped $La_2CuO_4$ cuprate which can be considered the archetypal high $T_c$ superconductor like $atomic$ $hydrogen$ in atomic physics.
In this work we discuss the variation of the superconducting dome changing the lattice geometry and the period of the superlattice made of N and S nanoscale layers shown in Fig.\ref{fig:1}. This is a selected prototype of an artificial superlattice of nanoscale quantum wells made of an interface space charge in a Mott insulator (LCO) layers confined by the normal the overdoped cuprate layers (LSCO) playing the role of potential barrier and charge reservoir shown in Fig.\ref{fig:1a}. In this heterostructure the quantum confinement of the space charge create two quantum subbands and their band width due to tunnelling of the electrons through the barrier is smaller of the energy cut-off of the pairing interaction. The particular case discussed in this work provides a archetypal example which grabs the essential physics for practical realisation of an artificial nanoscale heterostructure formed by the quantum superlattice with period in the transversal direction ranging between 3 and 5.28 nm.

\section{Theory and Methods}

The electron gas is confined in a superlattice of quantum wells  by a periodic potential along the z direction 
shown in  Fig. \ref{fig:1}, while in the xy plane it shows a free-electron-like energy dispersion. The non-interacting single-particle wave functions are given by
\begin{equation}
\psi_{n\mathbf{k}\lambda}\left(\mathbf{r}\right)=\varphi_{nk_{z}}\left(z\right)\frac{e^{i\mathbf{k}_{\parallel}\cdot\mathrm{r}_{\parallel}}}{\sqrt{\mathcal{A}}}\boldsymbol{\chi}_{\lambda}\left(\theta_{{\bf k}_{\parallel}} \right),\label{eq:wavefunction}
\end{equation}
with wave-vector components $\mathbf{k=}\left(k_{x},k_{y},k_{z}\right)\equiv\left(\mathbf{k}_{\parallel},k_{z}\right)$
label plane waves in the xy plane of area $\mathcal{A}$ and the Bloch
functions $\varphi_{nk_{z}}\left(z\right)$ along the z axis, $n$
being a subband index and $\lambda=\pm 1$: 
The helicity index. In the equation $\theta_{{\bf k}_{\parallel}}$ is defined as the angle between the $\mathbf{k}_{\parallel}$ wave-vector and the $k_x$ axis. The functions $\varphi_{nk_{z}}\left(z\right)$
and the corresponding eigenvalues are obtained by imposing the continuity of the wave function and its first derivative at the discontinuity points of the potential. ${\boldsymbol \chi}_{\lambda}(\theta_{{\bf k}_{\parallel}})$ are the spinors which are the eigenstates of the Rashba spin-orbit coupling. The pairing interaction among the electrons will be projected in the Rashba eigenstates [\onlinecite{gor2001superconducting}]. In the superlattices the quantum size effects split the electronic spectrum in $2$-subbands. 
In the two-gap superconductivity in our AHTS the Fano-Feshbach resonance occurs where the chemical potential is tuned near the band edge of the second ($n=2$) subband
[\onlinecite{valletta1997electronic,bianconi1998superconductivity,bianconi2005feshbach,innocenti2010resonant,innocenti2010shape,shanenko2012atypical,bianconi2014shape,jarlborg2016breakdown,mazziotti2017possible,cariglia2016shape,innocenti2010shape,innocenti2010resonant,guidini2014band,guidini2016bcs,perali1996gap,perali1997isotope,perali2004bcs,perali2004quantitative,perali2012anomalous,doria2016multigap,salasnich2019screening,shanenko2006size}] 
including the Rashba spin-orbit coupling [\onlinecite{gor2001superconducting,rashba1960properties,cappelluti2007topological,caprara2012intrinsic,caprara2014inhomogeneous,brosco2017anisotropy,mazziotti2018majorana,mazziottimultigap}].

The Rashba coupling constant [\onlinecite{mazziottimultigap}]  is given by :
\begin{equation}
\alpha=2\frac{\hbar^2}{2m} \frac{2\pi}{d} \alpha_{SO},
\end{equation} 
where $\alpha_{SO}$ is a dimensionless parameter describing the strength of the Rashba coupling in units of the 
modulation period of the superlattice $d$. In the numerical evaluation we use units such that the effective mass $m$ will be fixed to one. 
The unconventional  Lifshitz transition [\onlinecite{mazziottimultigap}] induced by spin-orbit coupling 
is associated with a particular unconventional van Hove 
singularity [\onlinecite{volovik2017lifshitz,volovik2017topological,volovik2018exotic}] 
in the negative helicity states.
The SOC breaks the spin degeneracy yielding two helicity states at fixed wave-vector $\mathbf{k}$) forming a circle whose radius increases with the SOC strength. 
The SOC shifts the bottom edge $E_L$ by an amount $E_0=-(m\alpha^2)/(2\hbar^2)$.
The Lifshitz parameter, defined in Eq.(\ref{eq:7.1}) takes into account the shift  by $E_0$ of the bottom edge $E_L$ in the presence of SOC.
The amplification of the superconducting gaps and $T_c$ appears when the chemical potential 
is tuned around the unconventional Lifshitz transition.
 The two-gap equations include the contact exchange interaction connecting pairs in different bands with different helicity of the pairing. The pair $\{(n, {\bf k}, \lambda),\ (n,- {\bf k}, \lambda)\}$ can be scattered into the pair  $\{(n', {\bf k}', \nu),\ (n',- {\bf k}', \nu)\}$, where $n$ and $n'$ are the band indices, $\textbf{k}$ and $\textbf{k}'$ are the wave-vectors and $\lambda$ and $\nu$ are the helicity indices. By recalling the results of our previous works (see Ref. [\onlinecite{mazziottimultigap}]), the zero-temperature two-gap equations read:
\begin{eqnarray}
\Delta_{\lambda,n} ({\mathbf k})&=&\lambda e^{i\theta_{{\mathbf k}_\parallel}} \Delta_n (k_z),\label{eq:11}\\ 
\Delta_n (k_z)&=&-\frac{U_0}{2}\sum'_{n',k'_z} I_{nk_z,n'k'_z} \Delta_{n'}(k'_z)  \sum'_{\nu,{\mathbf k'}_{\parallel}} \frac{1}{2E_{n', \nu, {\mathbf k'}}},
\nonumber
\end{eqnarray}
where the quasi-particle energy in the superconducting state, is
\begin{equation}
E_{n',\nu,{\mathbf k'}} =\sqrt{(\varepsilon_{\nu {\bf k'}_{\parallel}}+\varepsilon_{n'k'_z}-E_F)^2+|\Delta_{n'}(k'_z)|^2},
\label{eq:12}
\end{equation}
being understood that $\mathbf{k'}=\left(k'_{x},k'_{y},k_{z}\right)\equiv\left(\mathbf{k'}_{\parallel},k'_{z}\right)$.
The presence of SOC affects only the in-plane part of the wave functions. Hence, it is convenient to indicate the eigenvalues of the single-particle non-interacting Hamiltonian with the sum $\varepsilon_{\nu {\bf k'}_{\parallel}}+\varepsilon_{n'k'_z}$, where $\varepsilon_{\nu {\bf k'}_{\parallel}}={\hbar\bf k'}^2_{\parallel}/(2m)+\lambda \alpha { k'}_{\parallel}$ and $\varepsilon_{n'k'_z}$  is given by the numerical solution of the Kronig-Penney-like problem in the z direction.
The primed sums in Eq.(\ref{eq:12}) indicate that we consider only the pairs whose energies differ from the Fermi energy less than the renormalized energy cut-off, i.e. $|\varepsilon_{\nu {\bf k'}_{\parallel}}+\varepsilon_{n'k'_z}-E_F|<\omega_0$.
Finally, in Eq.(\ref{eq:12}) the important  exchange integral $I_{nk_z,n'k'_z}$, which carries the information of the motion along the z axis, reads
\begin{equation} 
I_{nk_z,n'k'_z}= \frac{2\pi}{L^2_\parallel}\int  |\varphi_{n,k_z}(z)|^2|\varphi_{n',k'_z}(z)|^2 dz.
\label{eq:13}
\end{equation}

In  Eq.(\ref{eq:11}) the pairing potential depends linearly on the helicity index $\lambda$ and through a phase factor on the in-plane momentum, whereas the quantity $\Delta_n(k_z)$ does not depend explicitly on both in-plane momentum and helicity [\onlinecite{gor2001superconducting}].
To obtain the critical temperature we need to consider the finite-temperature extension of the multi-gap equations of Eq.(\ref{eq:12}), which is 
\begin{equation}
\Delta_n(k_z)=-\frac{U_0}{2}\sum'_{n',k'_z} I_{nk_z,n'k'_z} \Delta_{n'}(k'_z)  \sum'_{\nu {\mathbf k'}_{\parallel}} \frac{\tanh (\frac{\beta E_{n',\nu,{\mathbf k'}}}{2})}{2E_{n',\nu,{\mathbf k'}}},
\label{eq:83}
\end{equation}
where $\beta=1/(k_BT)$. The critical temperature $T_c$ is obtained by taking the limit $T\rightarrow T_c$, $\Delta_n(k_z)$ approaching zero. More precisely, thanks to the matrix structure in the subband indices of the exchange integral, we get a linear system for the unknowns $\Delta_n(k_z)$, where we confine our analysis to $n=1$ and $n=2$. For $T=T_c$, the maximum eigenvalues of the matrix of the system is equal to 1 and this is the condition we seek when we solve the linearized gap equations for $T_c$. This matrix structure takes into account the interference between the electronic wave functions in the different subbands.
The possibility of varying each matrix term allows us to study the superconducting phase when different partial condensates coexist
in different coupling regimes and it amplifies the critical temperature reaching high-$T_c$ superconductivity
in the weak-coupling regime, with optimal choice of the nanoscale lattice parameters of a particular metallic heterostructure at the atomic limit [\onlinecite{mazziottimultigap},\onlinecite{mazziotti2017possible},\onlinecite{mazziotti2021resonant},\onlinecite{mazziotti2021room}].
 
 \begin{figure}
	\centering
	\includegraphics[scale=0.9]{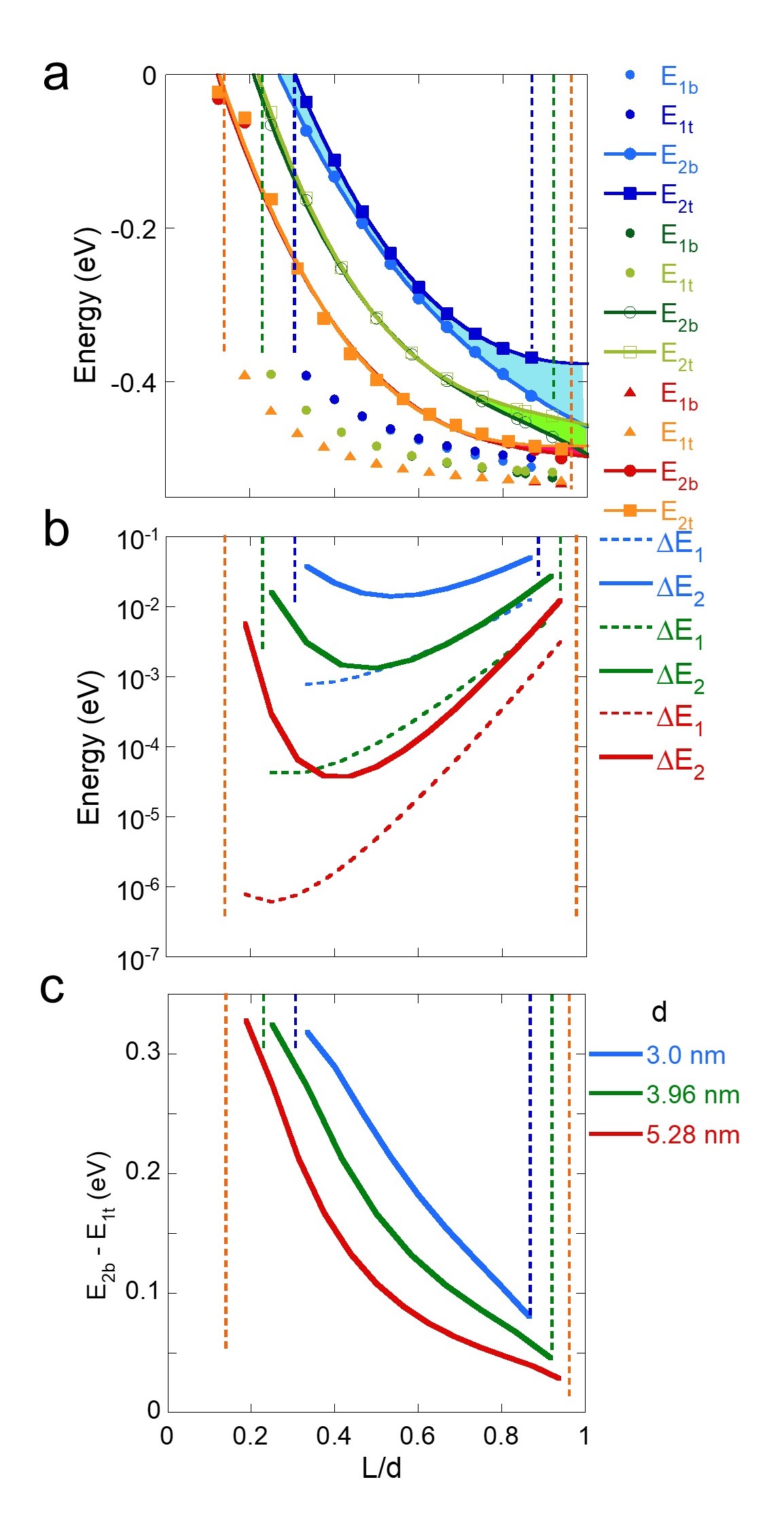} 
	\caption{ The energies of the second subband ${E_2}$ and of the first subband ${E_1}$ in panel (a),  
	the width of the second upper subband in panel (b), and the energy splitting between the two subbands $\Delta E=E_2-E_1$ in panel(c), 
	are plotted as function of the parameter $L/d$ of the geometry of the superlattice.
	}
\label{fig:2}
\end{figure}
 
\begin{figure}
	\centering
	\includegraphics[scale=0.9]{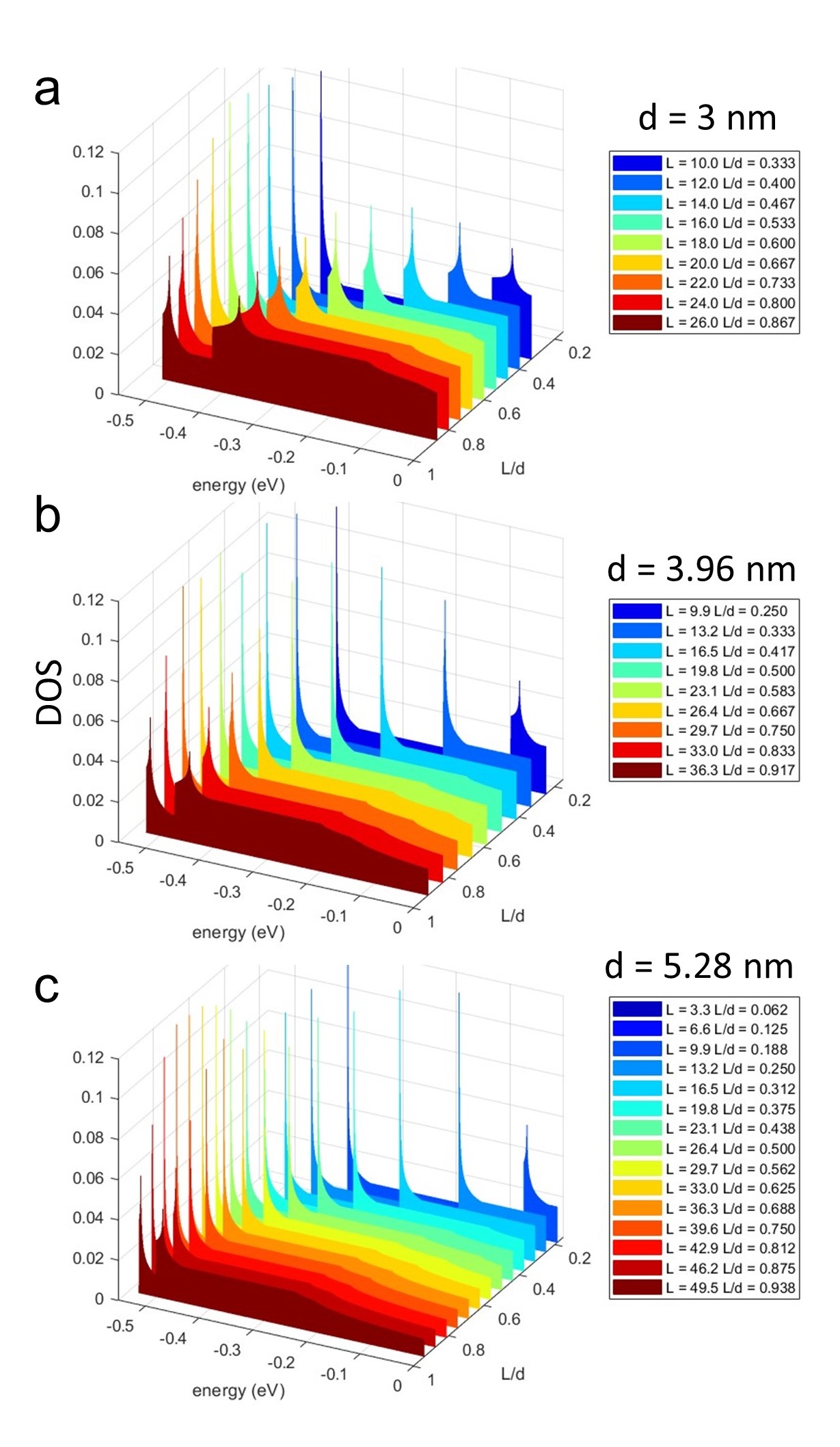}
	\caption{Panels (a), (b) and (c): total Density of States (DOS in arbitrary units) of the first two energy bands computed for three different periods d of the superlattice ((a): $d = 3 \ nm$, (b): $d = 3.96 \ nm$, (c): $d = 5.28 nm$) as a function of the Fermi energy, $E_F$, for different $L/d$.}
	  \label{fig:3}
\end{figure}

\begin{figure}
	\centering
	\includegraphics[scale=0.8]{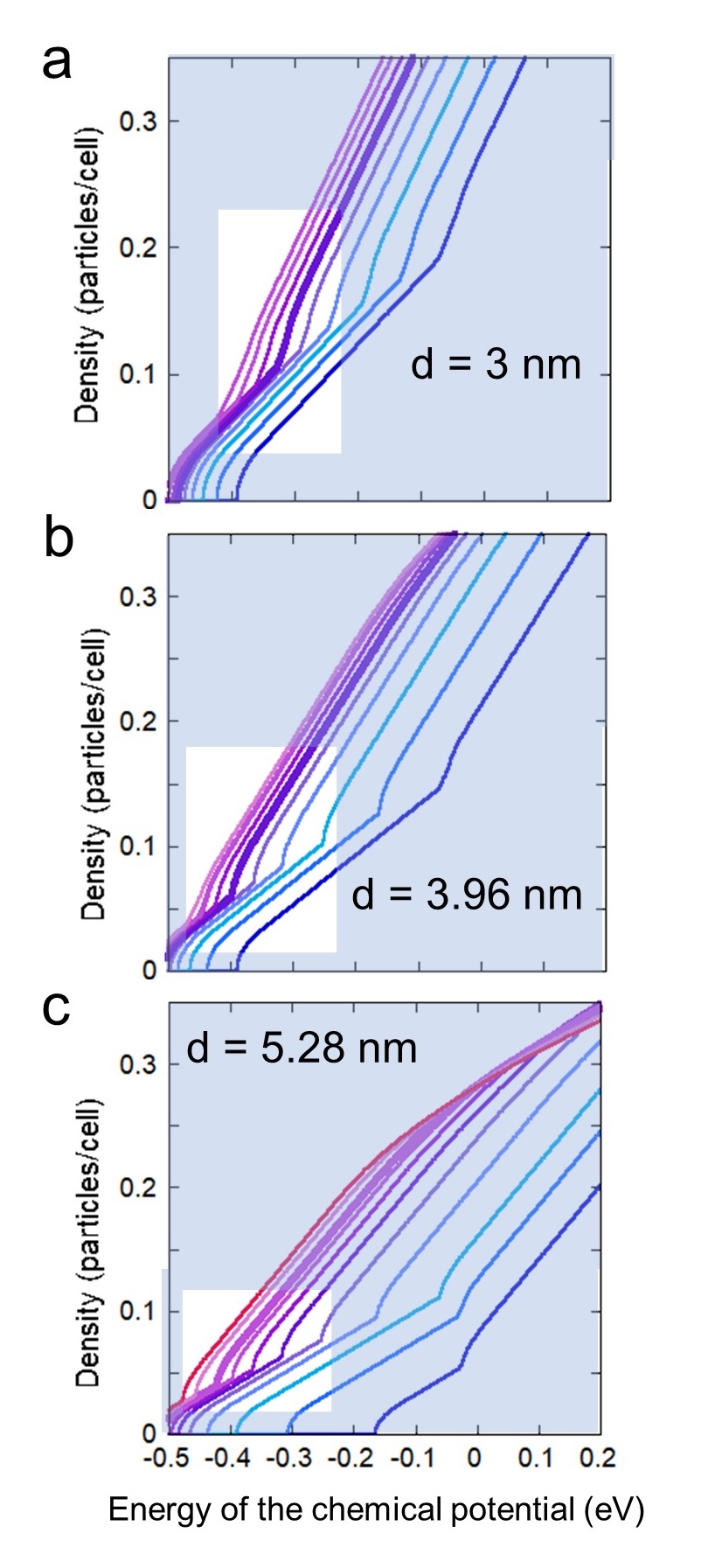}
	\caption{Panels (a), (b) and (c):  corresponding carrier density, $n$, in particles for unit cell,  computed for the three different $d$ used in panels (a), (b) and (c) respectively. The white rectangle evidences the regions where the torus to corrugated-to-cylinder electronic topological Lifshitz transition of the Fermi surface associated with a circular extended van Hove singularity occurs at a maximum $T_c$.}
	  \label{fig:4}
\end{figure}

\begin{figure}
	\centering
	\includegraphics[scale=0.35]{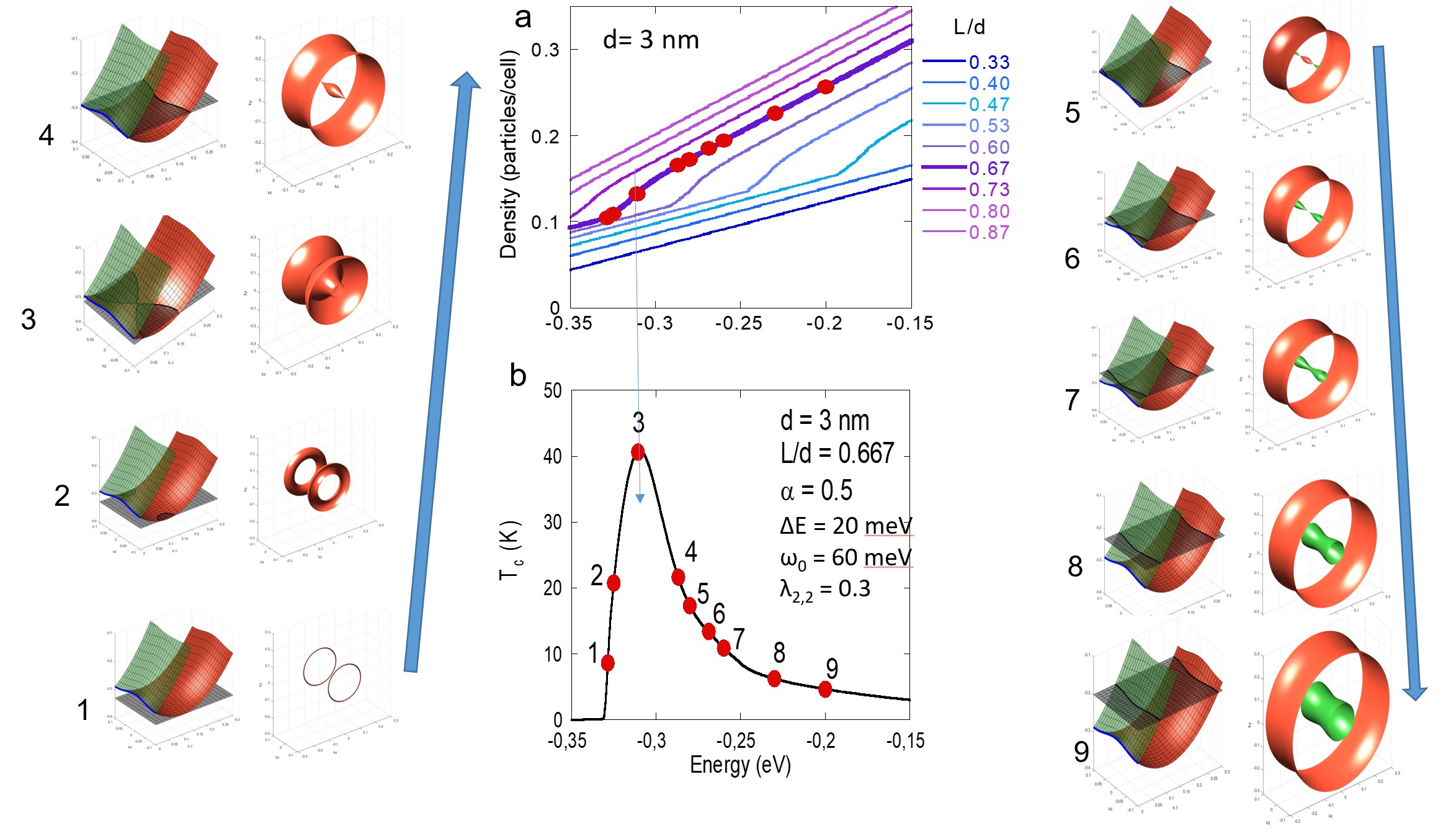} 
	\caption{Panels from (1) to (9): Representations of the Fermi surface for the superlattice with $d = 3\ nm$ and $L/d \approx 2/3$ computed for different Fermi energies with spin-orbit coupling. For each panel, in the subplot on the left the Fermi surface is reported as a function of $k_z$ and $k_{\parallel}$. The band energy dispersion along the z-direction is evidenced by a blue line at $k_{\parallel} = 0$. The positive and negative helicity sub-bands are indicated by the green and red surfaces, $\Lambda_+$ and $\Lambda_-$, respectively, while the gray plane at constant energy, $\Pi_{E_F}$, shows the values of the Fermi level for each panel. The intersection of $\Pi_{E_F}$ with $\Lambda_+$ and $\Lambda_-$ is evidenced by the black lines representing the Fermi surface projections in the $k_z - k_{\parallel}$ subspace. The Fermi surface in the full reciprocal space $(k_x, k_y, k_z)$ is shown in the right subplots and is obtained by rotating its projection on the $k_z - k_{\parallel}$ subspace, shown in the left panel, by 360 degrees along the $k_{\parallel} = 0$ line, which corresponds to the $(k_x = 0, k_y = 0)$ line of the right panel. The two branches of the Fermi surface corresponding to positive and negative helicity are colourful in green and red, respectively. The Fermi energies used in panels (1) - (9) are shown as red bullets in a detailed view of the $n$ vs $E_F$ plot in panel (a) and on the $T_c$ vs $E_F$ curve in panel (b).}
\label{fig:5}
\end{figure}

 \begin{figure}
	\centering
	\includegraphics[scale=0.55]{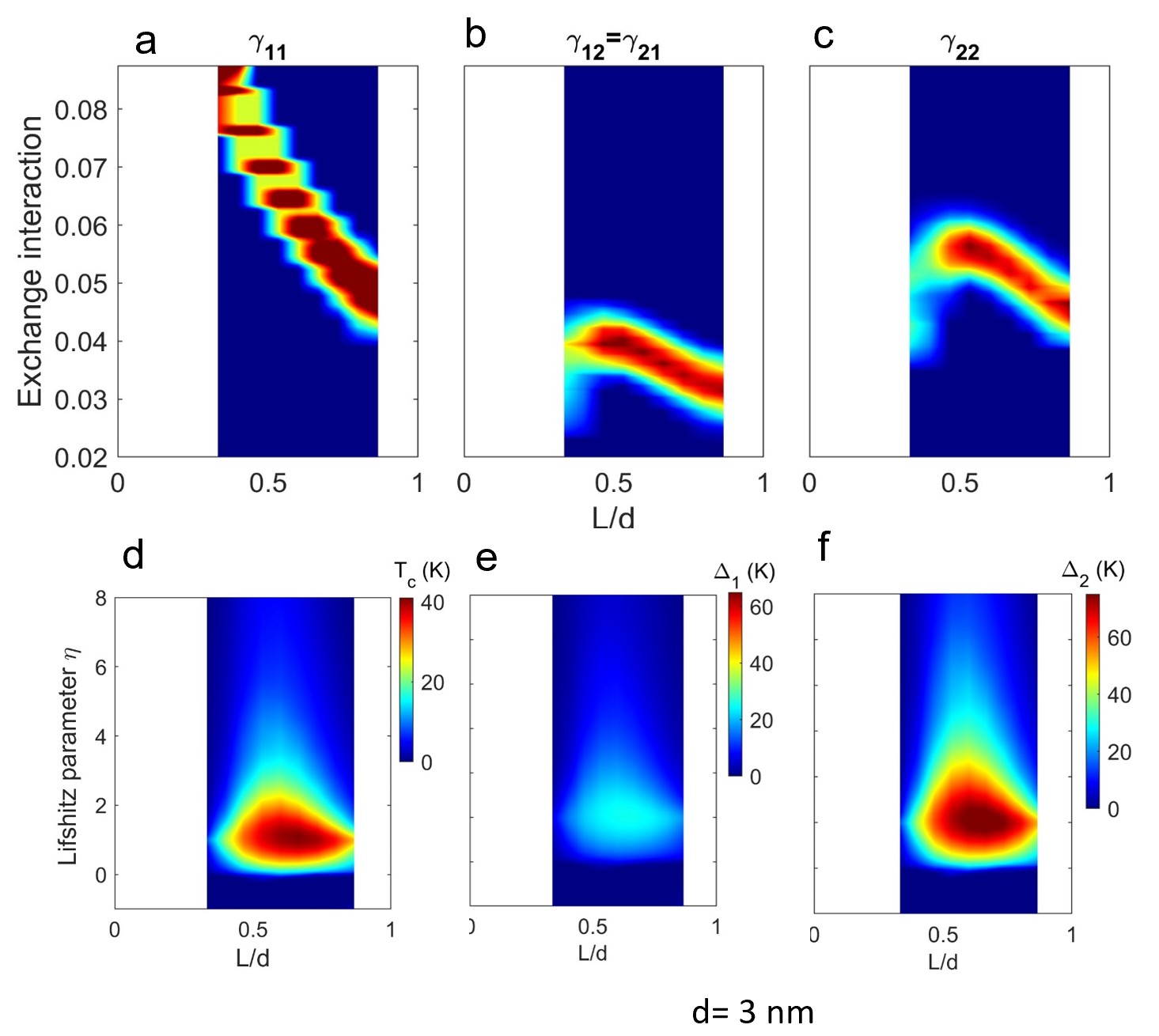} 
	\caption{The distribution functions of the exchange interactions ($\gamma_{11}$ in panel (a), $\gamma_{12} = \gamma_{21}$ in panel (b), and $\gamma_{22}$ in panel (c)) computed for the $d = 3\ nm$ superlattice as a function of the $L/d$ ratio. Panels (d), (e), and (f): critical temperature ($T_c$), first ($\Delta_1$) and second ($\Delta_2$) gap computed on the same superlattice as a function of the Lifshitz parameter and the $L/d$ ratio.}	
\label{fig:6}
\end{figure}

\begin{figure}
	\centering
	\includegraphics[scale=0.45]{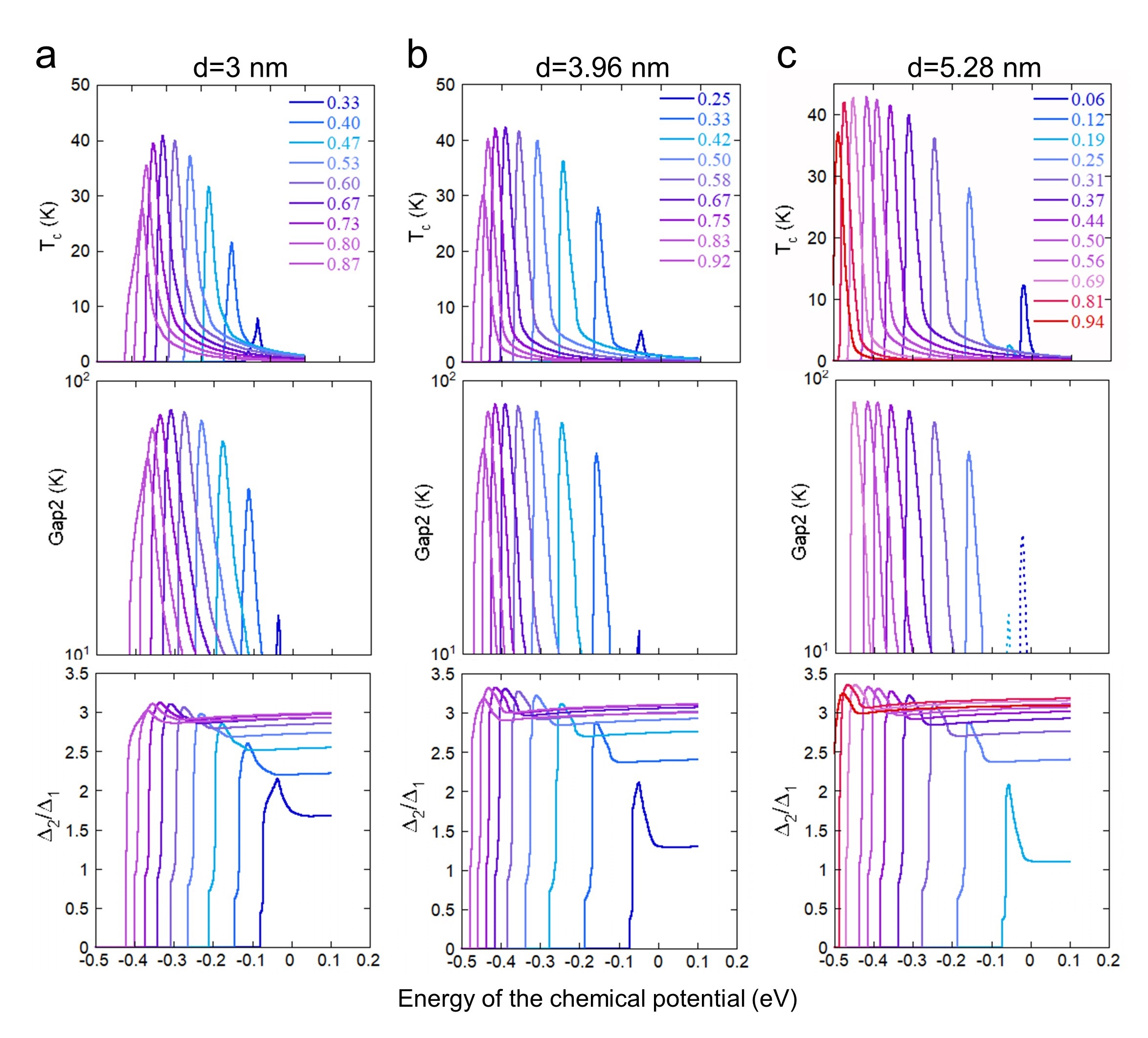} 
	\caption{Critical temperature ($T_c$  top row), second gap ($\Delta_2$, middle row) and second to first gap ratio ($\Delta_2/\Delta_1$, bottom row) calculated as a function of chemical potential energy for the superlattices with $d = 3\ nm$ (left column), $d = 3.96\ nm$ (middle column), $d = 5.28\ nm$ (right column) for different values of the $L/d$ ratio (see color legend at the top of each column).}
\label{fig:7}
\end{figure}

\begin{figure}
\centering
\includegraphics[scale=0.29]{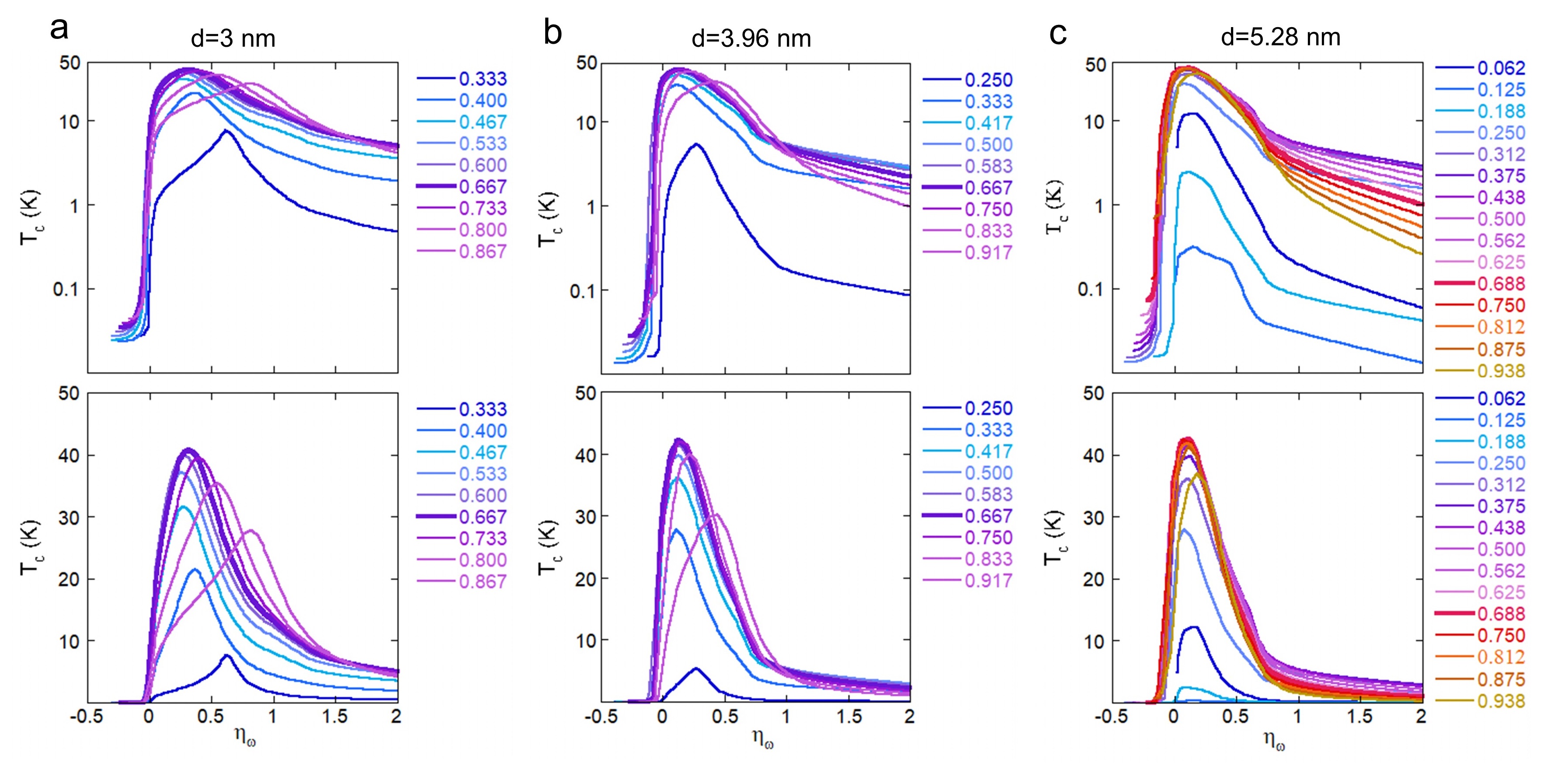}
\caption{Critical temperature ($T_c$) in linear (bottom-row) and logarithmic scale (top-row) computed as a function of the relative position of the Fermi energy with respect to the bottom of the second band in units of the cut-off energy ($\eta_{\omega}$). The left, middle and right columns show the results computed on the $d = 3\ nm,\ 3.96\ nm$ and $5.28\ nm$ superlattices, respectively, for different $L/d$.}
\label{fig:8}
\end{figure}

\section{Results and Discussion}

We start discussing the electronic properties of our superlattice heterostructure in its normal state showing the multi-band energy spectrum given by quantum size effect of the confined superconducting space charge confined in the modulation doped $La_2CuO_4$ quantum wells the density of states and corresponding Van Hove singularities, and the matrix elements of the multichannel pairing interaction.
Fig. \ref{fig:2} shows 
 (i) the energy of the second electronic subband ${E_2}$ and  of the first subband ${E_1}$, in panel (a),
 (ii) the width of the second upper subband, in panel (b) and 
 (iii) the energy splitting between the two subbands $\Delta E=E_2-E_1$, in panel (c),
 for superlattices with three different periods, ($d = 3 \ nm$,  $d = 3.96 \ nm$,  $d = 5.28 \ nm$)  
plotted as function of the parameter $L/d$ of the geometry of the superlattice.
The figure shows that where $L/d$ is less than 0.2-0.3 the second subband is pushed above the potential barrier and the system becomes a single band superconductor. 
On the contrary where  $L/d$ is larger than 0.8-0.9 the splitting between the two bands  vanishes and the system merge in an effective single band made of two overlapping bands. 
The two-gap superconductivity is possible only in the range $0.2<L/d<0.9$ 
 
 The total DOS of the first two energy bands computed for three different periods d of the superlattice of quantum wells ((a): $d = 3 \ nm$, (b): $d = 3.96 \ nm$, (c): $d = 5.28 nm$) are plotted In Fig. \ref{fig:3} as a function of the energy of the chemical potential and  the superlattice conformational parameter  $L/d$:  in the presence  of the spin-orbit coupling determined by the internal electric field at the interfaces for superlattices with three different periods panel (a) $d = 3 \ nm$, panel (b) $d = 3.96 \ nm$,  and panel (c) for $d = 5.28 \ nm$  
As it was shown in Ref.[\onlinecite{mazziottimultigap}], the peak in the DOS is linked to the topological phase transition of the Fermi surface. In particular, when the chemical potential reaches the Lifshitz transition with an unconventional extended VHS is observed where the singular nodal points on the Fermi surface form an entire circle, which explains the appearance of the sizable peak in the DOS. The radius of the circle  of singular points increases as the SOC increases and this is associated with the shift of the DOS peak to lower energy (see Ref.[\onlinecite{mazziottimultigap}]). 
The unconventional Lifshitz transition observed in the normal phase generates the amplification of the critical temperature. 
We used in this work a moderate spin-orbit coupling $\alpha_{SO}=0.4$ reaching  further increasing  up to $\alpha_{SO}=0.7$, it has been shown that it possible to reach at the top of the $T_c$ dome a critical temperature of $160\ K$, as in Hg-based cuprate perovskite superconductors. 

In order to obtain a comprehensive understanding of the rich behaviour of the normal state of the superlattice considered in this work, we have evaluated the carrier density and analysed the complex geometry and the topology of the Fermi surface determined by the SOC. The Fermi surface and its topology have been correlated with the superconducting state transition temperature.
Fig. \ref{fig:4} reports the carrier density, $n$, in particles per unit cell, as a function of the chemical potential computed for the three different values of $d$ used in panels (a), (b) and (c) respectively. The white rectangle evidences the regions where the torus to corrugated-to-cylinder electronic topological Lifshitz transition of the Fermi surface associated with a circular extended van Hove singularity occurs associated with  the variation of the superconducting critical temperature $T_c$.
Representations of the Fermi surface topology for the superlattice with period $d = 3\ nm$ and ratio $L/d \approx 2/3$ computed for different values of the  chemical potential in the presence of spin-orbit coupling are given in Fig. \ref{fig:5}. For each panel, in the subplot on the left the Fermi surface is reported as a function of $k_z$ and $k_{\parallel}$, which are the momentum component along the z-direction (see Fig.1) and the modulus of the momentum component parallel to the superlattice planes, respectively. The band energy dispersion along the z-direction is evidenced by a blue line at $k_{\parallel} = 0$. 
The positive and negative helicity subbands resulting from the SOC are indicated by the green and red surfaces, $\Lambda_+$ and $\Lambda_-$, respectively, while the gray plane at constant energy, $\Pi_{E_F}$, shows the values of the Fermi level chosen for each panel. The intersection of $\Pi_{E_F}$ with $\Lambda_+$ and $\Lambda_-$ is evidenced by the black lines representing the Fermi surface projections in the $k_z - k_{\parallel}$ subspace. The Fermi surface in the full reciprocal space $(k_x, k_y, k_z)$ is shown in the right subplots of each panel and is obtained by rotating its projection on the $k_z - k_{\parallel}$ subspace, shown in the left panel, by 360 degrees along the $k_{\parallel} = 0$ line, which corresponds to the $(k_x = 0, k_y = 0)$ line of the right panel. The two branches of the Fermi surface corresponding to positive and negative helicity are coloured in green and red, respectively. The Fermi energies used in panels (1) - (9) are shown as red bullets in a detailed view of the $n$ vs $E_F$ plot in panel (a) and on the $T_c$ vs $E_F$ curve in panel (b).

Having established the normal state properties, we investigate in the following the microscopic properties of the superconducting state of the superlattice heterostructure, using as input the electronic properties obtained in the normal state. 
First, we evaluate how the pairing interaction between electrons gets reconfigured by the geometric quantization of the single-particle wave-functions. Particular interest is assigned to the pair-exchange interactions, which is responsible for
the hybridization of the partial condensates forming from each subband and for the emerging behavior associated with the Fano-Feshbach resonance. With this information, we compute the most fundamental superconducting properties, as the superconducting energy gaps and the critical temperature of the super to normal state transition. Fig. \ref{fig:6} shows the distribution functions of the exchange interactions ($\gamma_{11}$ in panel (a), $\gamma_{12} = \gamma_{21}$ in panel (b), and $\gamma_{22}$ in panel (c)) computed for the $d = 3\ nm$ superlattice as a function of the $L/d$ ratio. Panels (d), (e), and (f): critical temperature ($T_c$), first ($\Delta_1$) and second ($\Delta_1$) gap computed on the same superlattice as a function of the Lifshitz parameter and the $L/d$ ratio.

\begin{figure}
\centering
\includegraphics[scale=0.35]{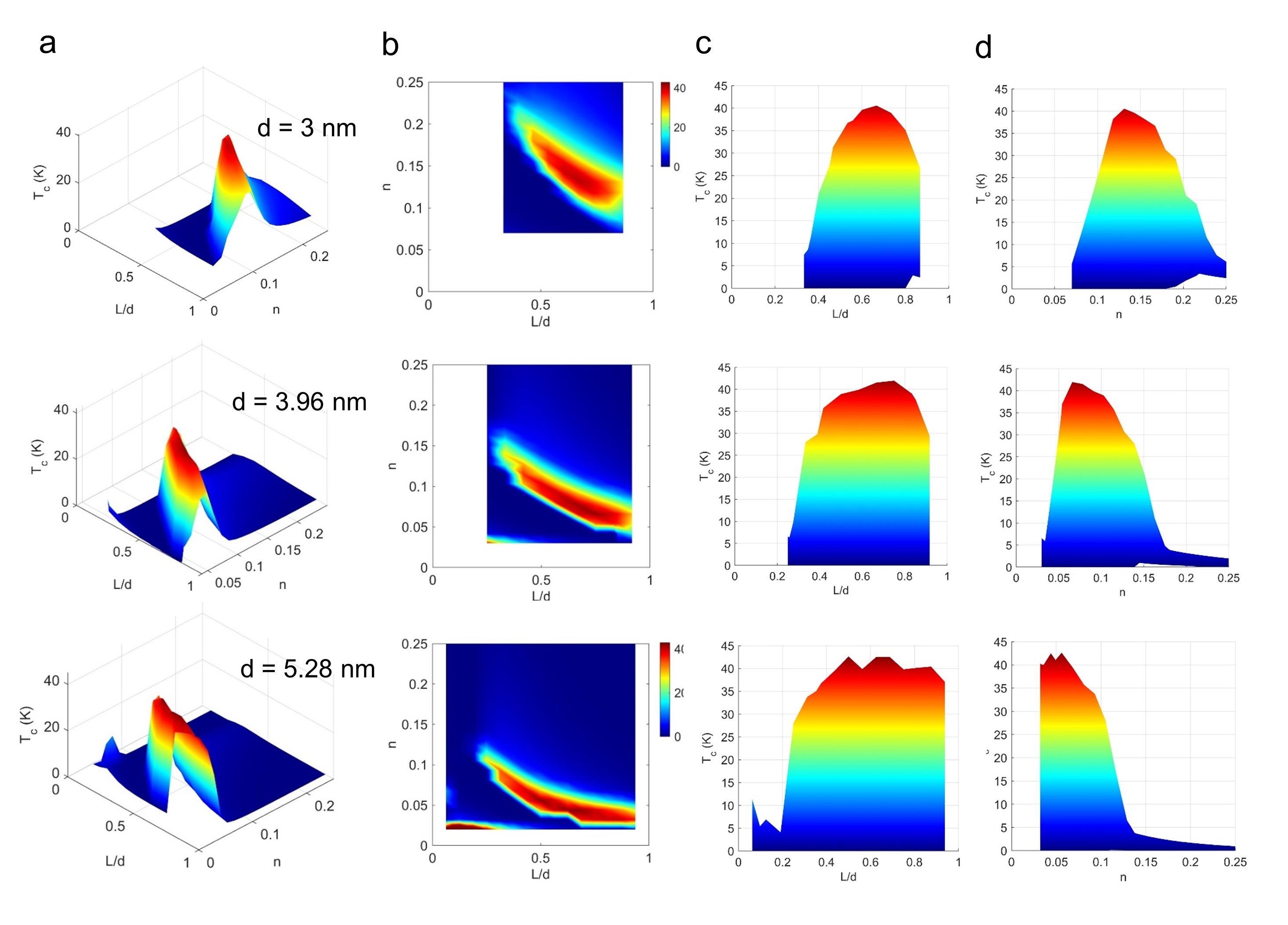}
\caption{Different views of $T_c$ as a function of density and $L/d$ for the superlattices with $d = 3\ nm$ (top row), $d = 3.96\ nm$ (middle row) and $d = 5.28\ nm$ (bottom row). From left to right: first column: 3D (isometric) view; second column: top view; third and fourth columns correspond to a projection on the $T_c$ vs $L/d$ plane  and a projection on the $T_c$ vs particle number $n$ plane.}
\label{fig:7}
\end{figure}

\begin{figure}
	\centering
	\includegraphics[scale=1.0]{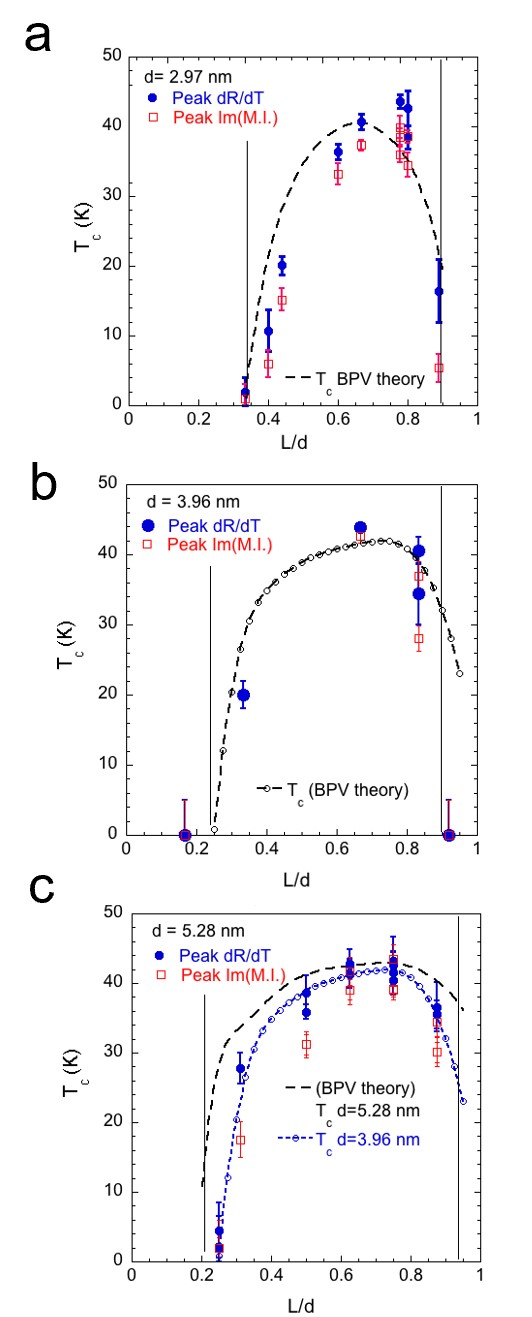}
	\caption{The theoretical $T_c$  (dashed line) in the superlattice with period $3\ nm$ (panel a) $3.96\ nm$ (panel b) and $5.28\ nm$ (panel c) as a function of $L/d$ calculated in this work are compared with the experimental data of $T_c$ versus $L/d$ for the artificial superlattice of superconducting $La_2CuO_4$ layers doped by interface space charge of thickness $L$ intercalated by $La_{1.55}Sr_{0.45}CuO_4$ layers of thickness W, with variable period d $2.97\ nm$, $3.96\ nm$ and $5.28\ nm$.}
	\label{fig:10}
\end{figure}

Critical temperature ($T_c$  top row), second gap ($\Delta_2$, middle row) and second to first gap ratio ($\Delta_2/\Delta_1$, bottom row) calculated as a function of chemical potential for the superlattices with $d = 3\ nm$ (left column), $d = 3.96\ nm$ (middle column), $d = 5.28\ nm$ (right column) for different values of the $L/d$ ratio are reported in Fig. \ref{fig:7}.
In Fig. \ref{fig:8} we report the $T_c$ in linear and logarithmic scale computed as a function of the relative position of the Fermi energy with respect to the bottom of the second band in units of the cut-off energy ($\eta_{\omega}$). The left, middle and right columns show the results computed on the $d = 3\ nm,\ 3.96\ nm$ and $5.28\ nm$ superlattices, respectively, for different values of the $L/d$ ratio. Different views of the calculated $T_c$ as a function of particle density and $L/d$ ratio for the superlattices with $d = 3\ nm$ (top row), $d = 3.96\ nm$ (middle row) and $d = 5.28\ nm$ (bottom row) are illustrated in Fig. \ref{fig:10}. From left to right: first column: 3D (isometric) view; second column: top view; third and fourth columns correspond to a projection on the $T_c$ vs $L/d$ plane (variable density case, to be compared with Fig.\ref{fig:8} and a projection on the $T_c$ vs $n$ plane, respectively.

\begin{figure}
	\centering
	\includegraphics[scale=1.0]{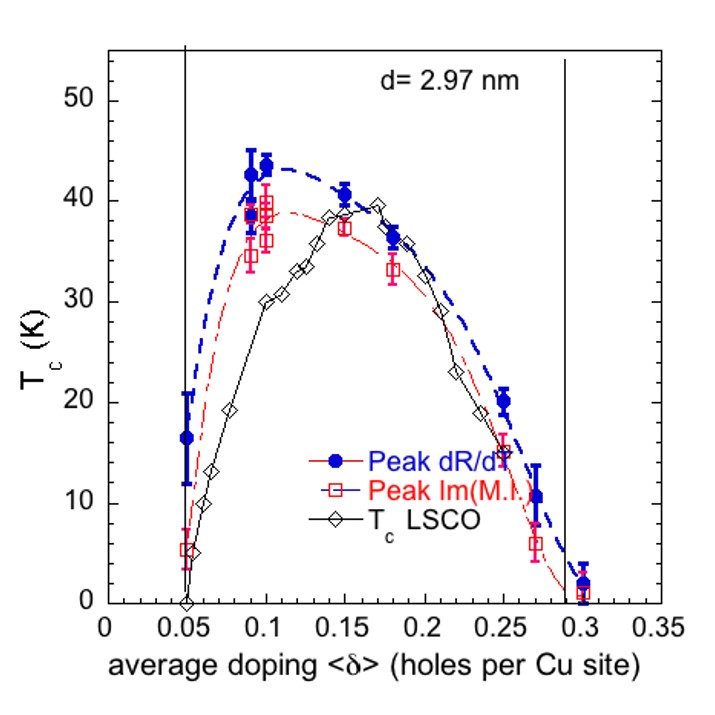}
	\caption{The theoretical $T_c$  (dashed line) in the artificial superlattice with period $3\ nm$ is plotted as function the electron number per unit volume given by the product of the of the charge density times the volume of a lattice unit of the $La_2CuO_4$ crystal. The experimental curves of $T_c$ versus doping (open black diamond points) in doped natural $La_{2-x}Sr_x CuO_4$ perovskite  [\onlinecite{takagi1989superconductor}] is plotted as a function doping . The experimental $T_c$  (solid blue dots measured by resistivity drop and red open squares measured by mutual inductance) in the artificial superlattice of superconducting $La_2CuO_4$ layers doped by interface space charge of thickness $L$ intercalated by $La_{1.55}Sr_{0.45}CuO_4$ layers of thickness $W$.}
	\label{fig:11}
\end{figure}

Once the theoretical framework has been clearly established, emerging from the BPV multichannel resonant theory applied to a breakthrough artificial heterostructure being able to explore the strong quantum confinement regime of a superconducting superlattice, we conclude our investigation by a direct quantitative comparison of the theoretical predictions obtained in this work with the available experimental data measured recently in the artificial oxide heterostructures and in the chemically sinthetised LaSrCuO doped cuprate superconductor. AHTS devices based on normal metal LSCO alternating with superconducting space charge layers in LCO thin layers have been synthesised via an ozone-assisted MBE method (DCA Instruments Oy) on $LaSrAlO_4$ (001) substrates. At the end of the procedure, the samples were cooled down in ozone  to $T_s$ = $493K$  At this temperature the stoichiometric $La_2CuO_4$ layers are free of oxygen interstitials dopants. Then the samples were cooled down in vacuum to avoid doping by oxygen interstitials  [\onlinecite{fratini2010scale,poccia2011evolution}].  The superconducting critical  temperature has a function of $L/d$ has been measured for each group of samples with the same superlattice period $d$ as recently described in [\onlinecite{logvenov2023superconducting}].

Fig.\ref{fig:10} shows the theoretical critical temperature (dashed line) in the superlattice of quantum wells with period $3\ nm$ (panel a) $3.96\ nm$ (panel b) and $5.28\ nm$ (panel c) as a function of $L/d$ calculated in this work, compared with the experimental data of $T_c$ versus $L/d$ for the artificial superlattice of superconducting $La_2CuO_4$ layers doped by interface space charge of thickness $L$ intercalated by $La_{1.55}Sr_{0.45}CuO_4$ layers of thickness W with experimental period $2.97\ nm$, $3.96\ nm$ and $5.28\ nm$ recently reported by Logvenov et al.  [\onlinecite{logvenov2023superconducting}].

The theoretical critical temperature (dashed line) in the artificial superlattice of quantum wells with period $3\ nm$ calculated in this work is plotted as function of the electron number per unit volume given by the product of the charge density times the volume of the lattice unit cell of the $La_2CuO_4$ crystal, see Fig. \ref{fig:11}. The experimental curves of $T_c$ versus doping (indicated by open black square points) in doped natural $La_{2-x}Sr_x CuO_4$ perovskite is plotted as a function doping.
The experimental critical temperature (solid blue dots measured by resistivity drop and red open squares measured by mutual inductance) in the artificial superlattice of superconducting $La_2CuO_4$ layers doped by interface space charge of thickness $L$ intercalated by $La_{1.55} Sr_{0.45}CuO_4$ layers of thickness $W$.

As we will highlight in the conclusions, our theoretical approach based on the BPV theory, implemented with spin-orbit coupling due to interface effects induced by phase separation, has a strong predictive power, being able to guide the quantum design at the nanoscale of artificial superconducting heterostructures, as well as to account for the high-$T_c$ dome of cuprate superconductors, circumventing  the strain dependent stripe phase giving a drop of $T_c$ at $x=1/8$ in the phase diagram of the natural oxide $La_{2-x}Sr_x CuO_4$ cuprate. [\onlinecite{bianconi2000stripe,agrestini2003strain,kuspert2024engineering}], which does not appear in AHTS samples. This finding points toward a universal theory for high-$T_c$ resonant superconductivity in systems having particular nanoscale geometry and multicomponent electronic structures, in which the different interacting channels generate Fano-Feshbach resonances shaping high-$T_c$ superconductivity. 

\section{Conclusions}

We have provided a quantum-geometric theory for engineering two-gap resonant superconductivity in cuprate artificial high-$T_c$ superlattices made of quantum wells by tuning the geometry of the superlattice with the period d ranging from 5.28 down to 3 nanometers. 
The interface internal electric field generating the Rashba spin-orbit coupling in the atomically thin superconducting heterostructure plays a key role in our theory. 
The $T_c$ and superconducting gaps result to be controlled by the Fano-Feshbach resonance between (i) the first superconducting gap in the first quasi-2D cylindrical Fermi surface and (ii) the second superconducting gap in the appearing Fermi surface at a Lifshitz ETT controlled by the SOC.  The maximum $T_c$ occurs at the ETT from a torus to a corrugated cylinder in the second subband, giving an extended  high-order VHS due to a finite circle of singular points. Our calculations demonstrate that high-$T_c$ superconductivity in artificial heterostructures with period with $3<d<5.28$ nm shows a universal maximum around $L/d=$2/3 and at the related optimal charge density range in the superlattice.

The quantitative comparison of our theoretical results with the experimental data shows that the theory can account for a number of striking features of the experimental results recently collected in artificial superlattices made of interface superconductivity in stoichiometric hole doped $La_2CuO_4$ layers of thickness L by electronic interface space charge, intercalated between normal metallic layers, $La_{1.55}$$Sr_{0.45}$$CuO_4$ of thickness W. The theoretical superconducting dome $T_c$ versus charge density predicted for the $d=3 nm$ superlattice reproduces the superconducting dome of $La_{2-x}$$Sr_{x}$$CuO_4$. Remarkably, the  $T_c$ drop assigned to pinned charge-density-waves around doping $x=0.12$ is not present, as indeed observed in recent experiments on artificial high-$T_c$ superlattices.

\begin{acknowledgments}
We acknowledge financial support of Superstripes onlus. This work has been partially supported by PNRR MUR project PE0000023-NQSTI. We acknowledge interesting discussions with Sergio Caprara, Giovanni Midei, and Roberto Raimondi. We are indebt to Vittoria Mazziotti for contributions in the early stage of this project. 
\end{acknowledgments}	

\bibliography{Bibliography} 

\end{document}